# Interstellar matter in Shapley-Ames elliptical galaxies.

## IV. A diffusely distributed component of dust and its effect on colour gradients

P. Goudfrooij[1,2] and T. de Jong[2,3]

[1] European Southern Observatory, Karl-Schwarzschild-Straße 2, D-85748 Garching bei München, Germany*
[2] Astronomical Institute "Anton Pannekoek", University of Amsterdam, Kruislaan 403, NL-1098 SJ Amsterdam, The Netherlands
[3] Laboratory for Space Research, P.O. Box 800, NL-9700 AV Groningen, The Netherlands*

**Abstract.** We have investigated *IRAS* far-infrared observations of a complete, blue magnitude limited sample of 56 elliptical galaxies selected from the Revised Shapley-Ames Catalog. Data from a homogeneous optical CCD imaging survey as well as published X-ray data from the *EINSTEIN* satellite are used to constrain the infrared data.

Dust masses as determined from the *IRAS* flux densities are found to be roughly an order of magnitude higher than those determined from optical extinction values of dust lanes and patches, in strong contrast with the situation in spiral galaxies. This "mass discrepancy" is found to be independent of the (apparent) inclination of the dust lanes. To resolve this dilemma we postulate that the majority of the dust in elliptical galaxies exists as a diffusely distributed component of dust which is undetectable at optical wavelengths.

Using *observed* radial optical surface brightness profiles, we have systematically investigated possible heating mechanisms for the dust within elliptical galaxies. We find that heating of the dust in elliptical galaxies by the interstellar radiation field is generally sufficient to account for the dust temperatures as indicated by the *IRAS* flux densities. Collisions of dust grains with hot electrons in elliptical galaxies which are embedded in a hot, X-ray-emitting gas is found to be another effective heating mechanism for the dust.

Employing model calculations which involve the transfer of stellar radiation in a spherical distribution of stars mixed with a diffuse distribution of dust, we show that the observed infrared luminosities imply total dust optical depths of the postulated diffusely distributed dust component in the range $0.1 \lesssim \tau_V \lesssim 0.7$ and radial colour gradients $0.03 \lesssim \Delta(B-I)/\Delta \log r \lesssim 0.25$. The observed *IRAS* flux densities can be reproduced *within the $1\,\sigma$ uncertainties in virtually all ellipticals in this sample* by this newly postulated dust component, diffusely distributed over the inner few kpc of the galaxies, and heated by optical photons and/or hot electrons. The radial colour gradients implied by the diffuse dust component are found to be smaller than or equal to the observed colour gradients. Thus, we argue that the effect of dust extinction should be taken seriously in the interpretation of colour gradients in elliptical galaxies.

We show that the amount of dust observed in luminous elliptical galaxies is generally higher than that expected from production by mass loss of stars within elliptical galaxies and destruction by sputtering in hot gas. This suggests that most of the dust in elliptical galaxies generally has an external origin.

## 1. Introduction

Although elliptical galaxies generally do not exhibit the large amounts of interstellar matter (ISM) typically found in spiral galaxies, it has recently become possible to detect various components of the ISM in elliptical galaxies (see, e.g., Roberts et al. 1991 and references therein; Goudfrooij et al. 1994b). These studies have shown that the presence of dust and gas in elliptical galaxies is the rule rather than the exception.

Besides detecting the interstellar matter, we are interested in its origin and fate since this material is expected to hold clues to the formation and subsequent evolution of elliptical galaxies through both the physical properties and the dynamics of the gas and dust. There are at least two possible sources of the observed ISM: accumulation of material shed by evolved, mass-losing giant stars within the galaxy (e.g., Faber & Gallagher 1976, Knapp et al. 1992), and accretion of gas during galaxy interactions. In order to systematically study the origin and fate of the ISM of elliptical galaxies we have conducted a deep, systematic optical survey of a complete, blue magnitude-limited sample of 56 elliptical galaxies drawn exclusively from the Revised Shapley-Ames Catalog of Bright Galaxies (hereafter RSA; Sandage & Tammann 1981). Deep CCD imaging has been performed through both broad-band filters and narrow-band filters isolating the nebular H$\alpha$+[N II] emission lines. The surface photometry and the isophotal properties of the sample galaxies are presented in Goudfrooij et al. (1994a, hereafter Paper I); optical images of dust and ionized gas are presented in Goudfrooij et al. (1994b, hereafter Paper II). Extinction properties of dust in selected dusty elliptical galaxies are presented in Goudfrooij et al. (1994c, hereafter Paper III), where dust masses are also estimated from the optical extinction data.

Optical CCD imaging is essential in establishing the presence and distribution of ISM in ellipticals in view of its high spatial resolution. However, quantitative estimates of the mass and the physical conditions of the dust from optical data alone are hampered by serious selection effects. This can be illustrated as follows: In Paper II, we reported an (optical) detection rate of dust features in our "RSA sample" (cf. Sect. 2) of 41%. However, the detectability of dusty disks or lanes in ellipticals by optical methods depends heavily on orientation effects, i.e., is biased towards edge-on dust distributions. The actual fraction of ellipticals containing dust features may thus be much higher (cf. Paper II; Sadler & Gerhard 1985). Furthermore, a significant fraction of the dust in elliptical galaxies may be expected to follow a smooth distribution similar to that of the stars (e.g., the dust which is produced by mass-loss from late-type stars within the galaxies), but this dust remains undetected by optical methods.

The means to study dust in galaxies in a more quantitative way has become available with data provided by the *InfraRed Astronomical Satellite (IRAS)*. The wavelength range observed by *IRAS* ($\lambda$ = 12, 25, 60, 100 $\mu$m) is crucial for studying dust in galaxies in a systematic way, for it is in this region that thermal radiation from dust with temperatures $300 \gtrsim T_d$ [K] $\gtrsim 20$ is emitted. Since the dust opacity is low in the infrared, orientation effects do not influence the detectability of dust, thus allowing quantitative estimates of the temperature and total mass of the dust. Using the technique of co-adding *IRAS* survey scans, Jura et al. (1987) and Knapp et al. (1989) have shown that a significant fraction ($\gtrsim 50\%$) of nearby, bright E and S0 galaxies have been detected at 60 and 100 $\mu$m at a limiting sensitivity about 3 times lower than in the *IRAS* Point Source Catalog. The detection rate by *IRAS* at the 60 and/or 100 $\mu$m passbands for our RSA sample is 61% (cf. Paper II). In addition to this significantly higher percentage with respect to the optical detection rate, recent studies of individual elliptical galaxies have shown that estimates of the total mass of dust using the *IRAS* data are a factor of $\sim$ 10 higher than those derived from optical colour excesses integrated over the areas where dust is found to reside (de Jong et al. 1990; Hansen et al. 1991; Goudfrooij et al. 1993). To address the discrepancies mentioned above, we combine in this paper our optical survey data with the *IRAS* data to analyse physical properties of dust in elliptical galaxies, including the distribution of the dust and heating mechanisms for the dust.

Following a short description of the galaxy sample (section 2), the far-infrared properties of elliptical galaxies are described in section 3. In Sect. 4 the optical and far-infrared data are combined to discuss the distribution of dust in elliptical galaxies. In Sect. 5 we discuss the origin and fate of dust in elliptical galaxies. A summary of the main conclusions is given in Sect. 6. This paper is based on part of the Ph. D. thesis of P. Goudfrooij which is available on request from him.

## 2. The galaxy sample

The elliptical galaxies discussed in this paper have been drawn from the (optically complete) sample consisting of all galaxies with $B_T^0 < 12$ denoted E in the RSA (cf. Paper I). This magnitude limit was chosen in accordance with the sample of Jura et al. (1987) to ensure the availability of coadded flux densities measured by *IRAS*, and coincides with the completeness limit of the RSA. The galaxies in the RSA sample are listed in Table 1 of Paper I, along with their assumed distances, magnitudes, sizes, and morphological classifications of both the RSA and the RC2 (de Vaucouleurs et al. 1976) catalogs. The absolute blue magnitudes of the members of this galaxy sample (hereafter called the "RSA sample") range between $-19.5$ and $-23.0$[1]. Hence, all galaxies of this sample belong to the class of "giant elliptical galaxies" according to the definition of Mihalas & Binney (1981).

## 3. Infrared properties of elliptical galaxies

We describe the infrared properties of the complete "RSA sample" of elliptical galaxies as deduced from the *IRAS* far-infrared observations in this Section.

The signature of far-infrared emission from spiral galaxies is the presence of strong long-wavelength radiation with typically $S(100) > S(60)$ (where $S(\lambda)$ is the *IRAS* flux density at wavelength $\lambda$). This is generally considered to be due to re-emission of dust heated by the ambient interstellar radiation field (e.g., de Jong et al. 1984) to an equilibrium temperature of $\sim 20$ K (Spitzer 1978; Draine & Anderson 1985). The *IRAS* data of bright early-type galaxies (Knapp et al. 1989) show that a significant fraction of these galaxies are detected at 60 and 100 $\mu$m, with flux density ratios similar to those of spiral galaxies. This suggests that the long-wavelength emission of elliptical galaxies is also due to emission by cool interstellar dust. However, other physical mechanisms might also be contributing significantly to the 60 and 100 $\mu$m flux densities, e.g., emission from (nonthermal) flat-spectrum nuclei, circumstellar dust surrounding late-type giant stars, or emission from a dusty torus near an active nucleus (if present). Goudfrooij (1994a) has assembled spectral energy distributions from the radio to the near-infrared spectral regions of all elliptical galaxies in the "RSA sample", and concluded that the 60 and 100 $\mu$m emission in these galaxies is virtually always due to cool interstellar dust (NGC 4261 and NGC 4494 were found to be possible exceptions to this "rule").

---

[1] $H_0 = 50$ km s$^{-1}$ Mpc$^{-1}$ is assumed throughout this paper

*3.1. Dust masses*

In this Section we derive dust masses from both far-infrared and optical data.

3.1.1. Dust masses from *IRAS* flux densities

Before deriving masses and luminosities of *cool* dust from the *IRAS* flux densities of the elliptical galaxies in our sample, we consider the emission of *hot* circumstellar dust in the 60 and 100 $\mu$m passbands. Assuming that circumstellar dust accounts for 40% of the 12 $\mu$m flux densities of elliptical galaxies (Knapp et al. 1992), we calculated the expected contribution to the 60 and 100 $\mu$m flux densities assuming diluted black body radiation (emissivity law $\propto \lambda^{-1}$) with a dust temperature of 350 K, appropriate to circumstellar dust surrounding M Mira stars in our Galaxy (e.g., Onaka et al. 1989). The resulting corrections to the 60 and 100 $\mu$m flux densities are:

$$S(60)_{\rm corr} = S(60) - 0.020\,S(12)$$
$$S(100)_{\rm corr} = S(100) - 0.005\,S(12)$$

In our sample, these corrections turn out to have little effect: the 60 $\mu$m flux densities are only affected by up to 2% (but usually much less).

The *IRAS* flux densities (corrected for the contribution of circumstellar dust) have been used to estimate the masses of cool dust. If $M_{\rm d}$ is the mass of dust, then

$$M_{\rm d} = \frac{D^2\,S_\nu}{\kappa_\nu\,B_\nu(T_{\rm d})} \qquad (1)$$

where $D$ is the distance, $S_\nu$ is the flux per unit frequency ("flux density"), $B_\nu(T_{\rm d})$ is the Planck function for the dust temperature $T_{\rm d}$ at frequency $\nu$, and $\kappa_\nu$ is the dust opacity,

$$\kappa_\nu = \frac{\pi a^2}{\frac{4}{3}\pi a^3 \rho_{\rm d}}\,Q_\nu$$

where $\rho_{\rm d}$ is the specific dust grain mass density, $Q_\nu$ is the grain emissivity factor, and $a$ is the average grain radius, weighted by its contribution to the far-infrared flux density and hence by the grain volume. For the grain size distribution function proposed by Mathis et al. (1977), we obtain $a = 0.1\ \mu$m. Adopting the values of $Q_\nu$ in the wavelength range 50 $\mu$m $\leq \lambda \leq$ 250 $\mu$m given by Hildebrand (1983) and $\rho_{\rm d} = 3$ g cm$^{-3}$, Eq. (1) becomes

$$M_{\rm d} = 5.1\,10^{-11}\,S_\nu\,D^2\,\lambda_\mu^4\,(e^{1.44\,10^4/\lambda_\mu\,T_{\rm d}} - 1)\ {\rm M}_\odot \qquad (2)$$

where $\lambda_\mu$ is in $\mu$m, $D$ is in Mpc, and $S_\nu$ is in mJy. The dust temperatures are put equal to the colour temperatures determined from the $S(100)/S(60)$ flux density ratios (corrected for the contribution of circumstellar dust) under the assumption that the far-infrared emission of elliptical galaxies originates from dust with an emissivity law $\propto \lambda^{-1}$ at wavelengths $\lambda \lesssim 200$ $\mu$m, typical of astronomical silicates (see, e.g., Schwartz 1982; Hildebrand 1983; Rowan-Robinson 1986; Mathis & Whiffen 1989; Kwan & Xie 1992). These temperatures should be regarded as

**Table 1.** Derived properties of the sample galaxies

| Galaxy | $\log L_B$ [L$_\odot$] | $\log L_{IR}$ [L$_\odot$] | $T_{\rm d}$ [K] | $\log M_{\rm d,IRAS}$ [M$_\odot$] | $\log M_{\rm d,opt}$ [M$_\odot$] |
|---|---|---|---|---|---|
| (1) | (2) | (3) | (4) | (5) | (6) |
| NGC 596  | 10.33 | < 8.41        | —        | < 6.09         | — |
| NGC 720  | 10.54 | < 8.17        | —        | < 5.87         | — |
| NGC 821  | 10.28 | $\leq$ 8.72   | $\leq$ 26 | 5.55 ± 0.10   | — |
| NGC 1395 | 10.36 | 8.45 ± 0.08   | 23 ± 6   | 5.60 ± 0.03    | — |
| NGC 1399 | 10.42 | $\leq$ 8.02   | $\leq$ 28 | 5.03 ± 0.12   | — |
| NGC 1404 | 10.31 | $\leq$ 8.09   | $\leq$ 27 | 4.98 ± 0.09   | — |
| NGC 1407 | 10.50 | 8.64 ± 0.06   | —        | 5.35 ± 0.27    | 3.48 |
| NGC 1427 | 9.96  | < 7.79        | —        | < 5.64         | — |
| NGC 1537 | 9.97  | $\leq$ 7.89   | $\leq$ 26 | 4.89 ± 0.12   | — |
| NGC 1549 | 10.19 | $\leq$ 7.72   | $\leq$ 29 | 4.55 ± 0.13   | — |
| NGC 1700 | 10.98 | < 9.03        | —        | < 6.93         | — |
| NGC 2300 | 10.41 | < 8.51        | —        | < 6.06         | — |
| NGC 2325 | 10.40 | < 8.70        | —        | < 6.32         | — |
| NGC 2974 | 10.29 | 9.27 ± 0.00   | 28 ± 1   | 6.20 ± 0.08    | 4.66 |
| NGC 2986 | 10.32 | $\leq$ 8.48   | $\leq$ 24 | 5.49 ± 0.13   | — |
| NGC 3136 | 10.22 | 8.29 ± 0.00   | $\geq$ 36 | 5.41 ± 0.09   | 4.78 |
| NGC 3193 | 9.91  | < 7.99        | —        | < 6.22         | — |
| NGC 3250 | 10.64 | < 8.88        | —        | < 6.41         | — |
| NGC 3377 | 9.70  | 7.82 ± 0.08   | 33 ± 8   | 4.28 ± 0.44    | 3.99 |
| NGC 3379 | 10.00 | < 7.67        | —        | < 5.28         | — |
| NGC 3557 | 10.86 | 9.34 ± 0.06   | 31 ± 5   | 5.94 ± 0.32    | — |
| NGC 3610 | 10.46 | < 8.54        | $\leq$ 28 | 5.34 ± 0.13   | — |
| NGC 3613 | 10.41 | < 8.38        | —        | < 6.06         | — |
| NGC 3640 | 10.25 | < 7.95        | —        | < 5.51         | 4.30 |
| NGC 3706 | 10.50 | 8.72 ± 0.00   | 30:      | 5.27 ± 0.67    | — |
| NGC 3904 | 10.16 | 8.76 ± 0.09   | 33 ± 8   | 5.21 ± 0.50    | — |
| NGC 3962 | 10.13 | 8.54 ± 0.18   | $\geq$ 32 | 5.49 ± 0.08   | 4.66 |
| NGC 4125 | 10.75 | 9.44 ± 0.16   | 35 ± 1   | 5.82 ± 0.08    | 5.65 |
| NGC 4261 | 10.56 | 8.50 ± 0.06   | 36 ± 17  | 4.64 ± 0.88    | — |
| NGC 4278 | 9.76  | 8.61 ± 0.00   | 31 ± 2   | 5.25 ± 0.09    | 4.37 |
| NGC 4365 | 10.28 | $\leq$ 8.29   | $\leq$ 25 | 5.16 ± 0.09   | — |
| NGC 4374 | 10.43 | 8.70 ± 0.17   | 35 ± 2   | 5.30 ± 0.12    | 4.54 |
| NGC 4373 | 10.71 | < 9.01        | —        | < 6.71         | — |
| IC 3370  | 10.69 | 9.85 ± 0.20   | 29 ± 1   | 6.73 ± 0.07    | 5.54 |
| NGC 4473 | 10.09 | < 7.93        | —        | < 5.55         | — |
| NGC 4486 | 10.67 | 8.52 ± 0.04   | 49 ± 9   | 4.07 ± 0.35    | 3.17 |
| NGC 4494 | 10.30 | $\leq$ 8.36   | $\geq$ 39 | 4.37 ± 0.11   | 3.92 |
| NGC 4564 | 9.77  | < 8.18        | —        | < 5.80         | — |
| NGC 4589 | 10.29 | 8.96 ± 0.00   | 31 ± 4   | 5.62 ± 0.30    | 4.93 |
| NGC 4621 | 10.25 | < 7.87        | —        | < 5.49         | — |
| NGC 4660 | 9.77  | < 7.90        | —        | < 5.52         | — |
| NGC 4697 | 10.46 | 8.86 ± 0.02   | 33 ± 2   | 5.21 ± 0.08    | — |
| NGC 4696 | 10.83 | 9.36 ± 0.00   | 24 ± 3   | 6.67 ± 0.17    | 5.65 |
| NGC 5018 | 10.63 | 9.75 ± 0.00   | 37 ± 2   | 5.95 ± 0.07    | 5.45 |
| NGC 5044 | 10.54 | 8.84 ± 0.00   | 48:      | 4.45 ± 0.46    | 4.26 |
| NGC 5061 | 10.41 | < 8.35        | —        | < 6.04         | — |
| IC 4296  | 11.00 | 9.20 ± 0.12   | 38:      | 5.36 ± 0.81    | — |
| NGC 5322 | 10.77 | 9.17 ± 0.03   | 35 ± 2   | 5.67 ± 0.11    | — |
| NGC 5576 | 10.27 | 8.34 ± 0.27   | 34:      | 5.41 ± 0.88    | 3.52: |
| NGC 5813 | 10.15 | < 8.16        | —        | < 6.02         | 3.98 |
| NGC 6482 | 10.96 | < 9.58        | —        | < 7.21         | 4.03 |
| NGC 7144 | 10.30 | 8.58 ± 0.11   | 30 ± 9   | 5.34 ± 0.65    | — |
| IC 1459  | 10.54 | 9.17 ± 0.02   | 35 ± 2   | 5.51 ± 0.11    | 5.27 |
| NGC 7507 | 10.33 | < 8.44        | —        | < 6.03         | 5.01 |

*Notes to Table 1.*
Column (2) lists the total blue luminosities of the galaxies, derived using a solar absolute magnitude M$_{\rm bol,\odot}$ = +4.75 (Allen 1973). The total infrared (i.e., 1- 500 $\mu$m) luminosities (cf. eq. (3) are listed in column (3). 2 $\sigma$ upper limits are given for the undetected galaxies. For the galaxies which are only detected at 60 or 100 $\mu$m, upper limits are denoted as '$\leq$'. Column (4) lists the dust temperatures as derived from the $S(100)/S(60)$ ratio. Lower or upper limits are calculated for galaxies which are only detected at 60 or 100 $\mu$m, respectively. Entries with a colon indicate uncertain values, due to marginal detections in both the 60 and 100 $\mu$m passbands. The dust masses estimated from the *IRAS* flux densities are listed in column (5). The entries for the galaxies which are detected *in one band only* are calculated using the upper or lower limit for the dust temperature as listed in column (4), and 2 $\sigma$ upper limits are given for the undetected galaxies for which a dust temperature of 30 K is assumed. Column (6) lists the dust masses as estimated from the optical extinction values, taken from Paper III.

"representative" values, since a *range* of temperatures is appropriate for dust within elliptical galaxies. Resulting values for $M_d$ and $T_d$ are listed in Table 1. We note that *IRAS* is sensitive to "cool" dust with $T_d \gtrsim 25$ K, but much less to "cold" dust with lower temperatures emitting predominantly at wavelengths far beyond 100 $\mu$m (e.g., Young et al. 1986; van den Broek et al. 1991). Since temperatures of order 20 K are appropriate to the outer regions of elliptical galaxies (i.e., beyond $\sim 10$ kpc, cf. Sect. 4.1; Jura 1982), the dust masses calculated by means of Eq. (2) are in principle lower limits to the real dust mass. In the following, we thus only consider the component of dust which radiates at wavelengths covered by *IRAS*.

### 3.1.2. Dust masses derived from extinction at optical wavelengths

In Papers II and III we derived dust masses from the optical (total) extinction values integrated over the galaxy areas where dust is found to reside. These optically derived dust masses are also listed in Table 1 to enable a quantitative comparison of the different derivations of the dust mass (cf. Sect. 4).

### 3.2. Far-infrared luminosities

Total (far-)infrared (i.e., 1–500 $\mu$m) luminosities of interstellar dust in the elliptical galaxies in our sample have been calculated using the 60 and 100 $\mu$m *IRAS* flux densities (corrected for the contribution of circumstellar dust), following the method described in the "Cataloged Galaxies and Quasars in the *IRAS* Survey" (JISWG 1986). Assuming a single temperature component of dust and a emissivity law $\propto \lambda^{-1}$, the infrared luminosity ($L_{IR}$) is given by

$$L_{IR} = 3.75 \, 10^3 \, D^2 \, C \, (2.58 \, S(60) + S(100)) \, L_\odot \qquad (3)$$

where $D$ is the distance in Mpc, $S(60)$ and $S(100)$ are *IRAS* flux densities in mJy, and the constant $C$ corrects for the flux missed shortward of 40 $\mu$m and beyond 120 $\mu$m and is a function of the dust temperature (i.e., the $S(100)/S(60)$ ratio). The values of C are listed in Table B.1 of the "Cataloged Galaxies and Quasars in the *IRAS* Survey" (JISWG 1986). The calculated values of $L_{IR}$ are listed in Table 1.

## 4. The distribution of dust in elliptical galaxies

In this Section we will combine our optical imaging data from Paper II and Paper III with the available *IRAS* data to discuss the distribution of dust in elliptical galaxies.

A glance at Table 1 reveals that the dust masses estimated from the optical extinction are significantly *lower* than those estimated from the far-infrared emission. Quantitatively, the average ratio $<M_{d,\mathrm{IRAS}}/M_{d,\mathrm{opt}}> = 8.4 \pm 1.3$ for the galaxies in Table 1 for which the presence of dust is revealed by both far-infrared emission and optical dust lanes or patches. Strictly speaking, the dust masses derived from optical extinction are lower limits since the conversion factor from extinction to dust mass implicitly assumes the dust to be in front of the stars.

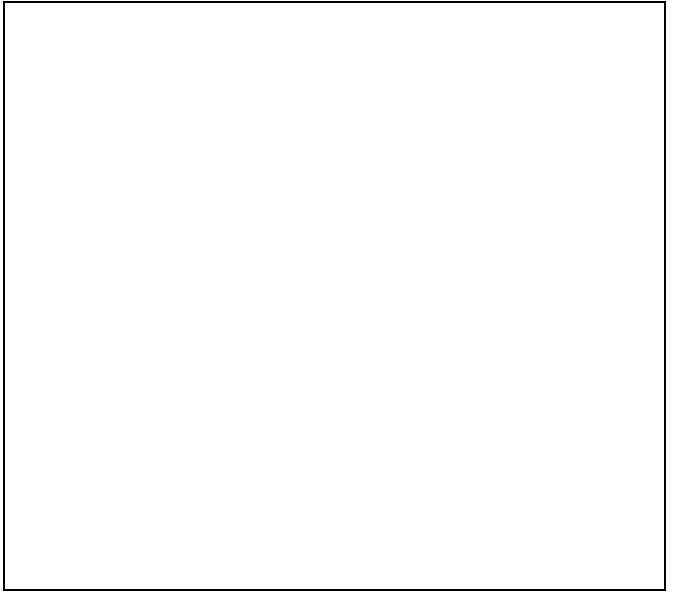

**Fig. 1.** The quotient of the dust mass derived from the *IRAS* data and the dust mass derived from optical extinction versus the cosine of the inclination angle of the dust lane in selected elliptical galaxies containing regular dust lanes. Filled squares represent galaxies from the "RSA sample", filled circles represent galaxies from Paper III, and open squares represent galaxies from the sample of Véron-Cetty & Véron (1988). The NGC numbers of the galaxies involved are indicated above the symbols

Since the dust is probably embedded in the stellar systems, the optically derived dust masses may be of order 2 times higher with respect to the tabulated values. However, the disagreement between the the two mass determinations is significant.

This disagreement might (partly) be due to the effect of orientation. In Paper II we reported that estimated inclinations of regular dust lanes in early-type galaxies are found to range between 0 and $\sim 35°$, suggesting that dust lanes are detected only if inclined by less than a certain critical angle (see also Sadler & Gerhard 1985). If this were correct, the ratio $M_{d,\mathrm{IRAS}}/M_{d,\mathrm{opt}}$ (hereafter "dust mass discrepancy") would be expected to be inversely proportional to $\cos(i)$, where $i$ is the inclination of the dust lane with respect to the line of sight. To evaluate the influence of the effect of orientation on the dust mass discrepancy, we estimated the inclinations of (apparently) regular, uniform dust lanes in elliptical galaxies from images shown in homogeneous optical CCD surveys (Paper II; Paper III; Véron-Cetty & Véron 1988). The dust masses derived from the optical reddening by Véron-Cetty & Véron were scaled up by 10% proportional to the difference of the adopted dust mass column densities (cf. Paper III), and values of $M_{d,\mathrm{IRAS}}$ for the galaxies taken from Véron-Cetty & Véron were calculated using Eq. (2). The relation between $M_{d,\mathrm{IRAS}}/M_{d,\mathrm{opt}}$ and $\cos(i)$ is shown in Fig. 1. This being a scatter plot, the effect of orientation on the dust mass discrepancy must be weak if present at all. This suggests that the dust in the lanes is concentrated in relatively small, dense clumps with a low volume filling factor.

Another argument in favour of the insignificance of orientation effects is provided by the observation that the distribution of axial ratio (log a/b) for the elliptical galaxies from the RSA sample in which the presence of dust is revealed through both dust lanes and far-infrared emission is indistinguishable with that for the ellipticals with *only* far-infrared emission.

In the case of axisymmetric (i.e., oblate or prolate) galaxies, the probability of detecting a dust lane would be expected to be lowest if the galaxy is very close to face-on, i.e., at the lowest ellipticities. Having eliminated the effect of orientation, the most plausible way out of the dilemma of the dust mass discrepancy is to postulate an additional component of dust with a diffuse, uniform distribution over a large area in elliptical galaxies and therefore virtually undetectable by optical methods (cf. also de Jong et al. 1990; Hansen et al. 1991; Goudfrooij et al. 1993). We note that this diffuse component of dust is not unexpected in elliptical galaxies. E.g., the late-type stellar population of typical giant ellipticals ($L_B = 10^{10} - 10^{11}$ L$_\odot$) has a substantial present-day mass loss rate ($\sim 0.1 - 1$ M$_\odot$ yr$^{-1}$ of gas and dust; cf. Faber & Gallagher 1976; Knapp et al. 1992) which is expected to be diffusely distributed.

An interesting potential way to trace this diffuse component of dust is provided by radial colour gradients in elliptical galaxies. With very few significant exceptions, "normal" elliptical galaxies (e.g., the galaxies in our RSA sample) show a reddening toward their centres (see Paper I and references therein). Gradients in one colour correlate well with gradients in other colours from the ultraviolet through the near-infrared (Peletier et al. 1990b). The colour gradients are generally being explained in terms of metallicity variations. This view is supported by the observation of radial gradients of metallic absorption line strengths (Kormendy & Djorgovski 1989; Davies et al. 1993; Carollo et al. 1993). There is however significant scatter in the relation between colour gradients and line strength gradients (cf. Peletier 1989; Davies et al. 1993). Although the presence of dust in elliptical galaxies is now beyond dispute, the implications of dust extinction are generally discarded in the interpretation of colour gradients. However, recent model calculations of the transfer of stellar radiation within elliptical galaxies by Witt et al. (1992, hereafter WTC) have demonstrated that a diffuse distribution of dust throughout ellipticals can cause significant colour gradients even with modest dust optical depths, without significantly changing the observed global colours.

We have exploited the elliptical galaxy model described by WTC to derive the predicted colour gradients – due to dust alone – appropriate to the far-infrared properties (and hence the dust content) of the elliptical galaxies in the RSA sample. The Monte Carlo simulations of WTC exploit a spherically symmetric distribution of stars, uniformly mixed with a diffuse, homogeneous distribution of dust. The predicted radial colour gradients are due to the effect of differential extinction (i.e., absorption *and* scattering) throughout the galaxy. The assumed density law of the dust distribution decreases with radius as $\rho_{\rm dust} \propto r^{-1}$, which generates radial colour gradients that are linear in log (radius), compatible with observed colour gradients

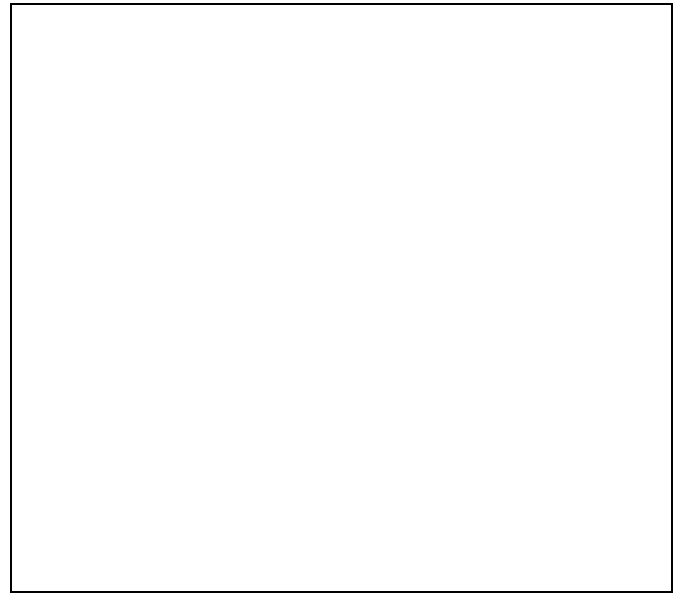

**Fig. 2.** The relation of $L_{IR}/L_B$ with the radial $B - I$ colour gradients (defined as $\Delta (B - I)/\Delta (\log r)$) for elliptical galaxies in the "RSA sample". Filled squares represent galaxies detected by *IRAS* showing optical evidence for dust, and open squares represent galaxies detected by *IRAS* without optical evidence for dust. Arrows pointing downwards indicate upper limits to $L_{IR}/L_B$. The dotted line represents the colour gradient expected from differential extinction by a diffuse distribution of dust (see text)

in elliptical galaxies (see Paper I and references therein; Wise & Silva 1994). The model of WTC reproduces a de Vaucouleurs (1959) surface brightness profile quite well, even for relatively high total optical depths of the dust distribution. Substantial amounts of dust can thus be present in (elliptical) galaxies while remaining undetected by conventional optical methods (see also Puxley & Brand 1994 for other examples of this effect). Using the model computations of WTC, infrared to optical luminosity ratios $L_{IR}/L_{\rm opt}$ and colour gradients [e.g., $\Delta (B - I)/\Delta (\log r)$] have been derived as a function of the total dust optical depth $\tau_V$ of the diffuse dust component. The conversion of $L_B$ to $L_{\rm opt}$ used by Thronson et al. (1990) was adopted.

The result is plotted (as a dotted line) in Fig. 2 in which the $L_{IR}/L_B$ ratios of the galaxies in the RSA sample that are detected by *IRAS* are plotted versus their radial $B - I$ colour gradients (taken from Paper I). Although Fig. 2 shows that colour gradients in elliptical galaxies are generally larger than can be generated by a diffuse distribution of dust throughout the galaxies according to the model of WTC it is obvious that the relation implied by the model of WTC fits the *minimum* of the distribution of colour gradients at a given $L_{IR}/L_B$ quite well. Moreover, there are *no* galaxies in the RSA sample with a colour gradient significantly *smaller* than that indicated by the model of WTC. This result strongly suggests the presence of a diffuse distribution of dust in elliptical galaxies, causing a colour gradient due to differential extinction (see also Wise & Silva 1994). This effect should be added to the effects of metallicity and/or age gradients of the stellar population in the interpretation of observed colour gradients of elliptical galaxies.

We have investigated whether published radial metallicity gradients (i.e., radial gradients of the Mg$_2$ index) of galaxies in the RSA sample are consistent with the positions of the galaxies in Fig. 2. Although there are currently only a few galaxies in the RSA sample for which radial metallicity gradients have been published, it is quite reassuring that the galaxies in Fig. 2 which lie nearest to the relation implied by the model of WTC have metallicity gradients which are among the smallest of the published values (e.g., NGC 3136, log d(Mg$_2$)/dr $= -0.035 \pm 0.011$; IC 3370, log d(Mg$_2$)/dr $= -0.026 \pm 0.010$ (Carollo et al. 1993)).

The consistency of the postulated diffuse distribution of dust in elliptical galaxies in terms of total masses and energetics of the dust will be investigated below.

*4.1. Radial dust temperature distribution*

To check whether the assumption of the presence of a diffuse, uniformly distributed component of dust in elliptical galaxies is energetically consistent with the available *IRAS* data, we investigate plausible heating mechanisms for the dust in this Section.

*Heating by stellar photons*
First we consider heating by starlight. De Jong (1986) and Jura (1986) have proposed that the far-infrared radiation of elliptical galaxies can be explained by thermal emission of dust grains heated by the general interstellar radiation field. This heating mechanism will be systematically investigated here using the extensive surface photometry of our sample (cf. Paper I).

Defining $I_{\rm opt}(r)$ as the average intensity of optical radiation (in units of erg cm$^{-2}$ s$^{-1}$ ster$^{-1}$) at galactic radius $r$, the heating rate $H_{\rm opt}(r)$ of a dust grain by absorption of optical light can be written

$$H_{\rm opt}(r) = \langle 4\pi\, I_{\rm opt}(r)\, Q_{\rm abs, opt} \rangle \pi a^2 \quad {\rm erg\ s}^{-1}, \qquad (4)$$

whereby $a$ is the (weighted mean) grain radius. The "average" value of $I_{\rm opt}(r)\, Q_{\rm abs, opt}$ is weighted over the wavelength range $0.25\ \mu{\rm m} \leq \lambda \leq 2.25\ \mu{\rm m}$ using the the characteristic spectral energy distribution for early-type galaxies given by Guiderdoni & Rocca-Volmerange (1987), and the $Q_{\rm abs}$ values for graphite and "dirty silicate" grains given by Jones & Merrill (1976). We assumed equal abundances of graphite and silicate dust grains. This weighting procedure results in

$$\langle I_{\rm opt}(r)\, Q_{\rm abs, opt} \rangle = 1.1\, I_{\rm V}(r)\, Q_{\rm abs, V}$$

The V band intensities are derived from the radial surface brightness profiles given in Paper I, assuming that V = 0.0 corresponds to $3.5\,10^{-20}$ erg s$^{-1}$ cm$^{-2}$ Hz$^{-1}$ (Hayes & Latham 1975).

*Heating by hot electrons in X-ray-emitting gas*
Observations with the *EINSTEIN* X-ray satellite have shown that a substantial number of elliptical galaxies in our sample are embedded in a hot (T $\sim 10^7$ K), X-ray-emitting gas (e.g., Forman et al. 1985; Fabbiano et al. 1992). The presence of this hot gas has important implications for the dust grains. Collisions of dust grains with protons and $\alpha$-particles in the hot gas destroy the dust by sputtering (Barlow 1978; Draine & Salpeter 1979a) while collisions with "hot" electrons provide a potentially important heating mechanism for the dust grains (Dwek 1986; de Jong et al. 1990).

The *EINSTEIN* satellite has provided us with data of sufficient quality to allow detailed studies of the radial distribution of the surface brightness and temperature of the gas for about 10 early-type galaxies (Forman et al. 1985; Trinchieri et al. 1986). The fitted average gas temperatures are in the range $0.5\,10^7$ K $\leq T \leq 4\,10^7$ K (Kim et al. 1992). The measured surface brightness profiles $\Sigma(r)$ are generally consistent with a "King-like" distribution of the form

$$\Sigma(r) = \Sigma(0) \left[ 1 + \left( \frac{r}{a_{\rm x}} \right)^2 \right]^{-3\beta + 1/2} \qquad (5)$$

(Cavaliere & Fusco-Femiano 1976) with $\beta \simeq 0.45$ (Forman et al. 1985; Trinchieri et al. 1986), where $r$ is the radial distance from the galaxy centre and $a_{\rm x}$ is the (X-ray) core radius. In the case of isothermal gas, Eq. (5) corresponds to an electron density profile of

$$n_{\rm e}(r) = n_{\rm e}(0) \left[ 1 + \left( \frac{r}{a_{\rm x}} \right)^2 \right]^{-3\beta/2}. \qquad (6)$$

Hence, core radii determined by fitting the X-ray surface brightness distribution to Eq. (5) can be used to compute electron density profiles for the gas if the gas is isothermal. The X-ray core radius has been measured for only a few ellipticals (Forman et al. 1985; Trinchieri et al. 1986; Killeen & Bicknell 1988). Most measurements indicate a core radius $a_{\rm x} \sim 2$ kpc (using H$_0$ = 50 km s$^{-1}$ Mpc$^{-1}$), with no clear correlation or systematic trend with optical luminosity. The high angular resolution measurements of Trinchieri et al. yielded somewhat smaller core radii; however, cooling processes of the gas can produce a positive temperature gradient (i.e., dT/dr > 0) in the central region of the galaxies (see, e.g., the review of Forman 1988). Since the X-ray emissivity increases with decreasing temperature over the relevant temperature range (McKee & Cowie 1977), this effect would produce spuriously small core radii and, hence, high central electron densities. In the following we will therefore use the core radii which have been fitted to the outer regions of the surface brightness distribution. Using the calibration given by Canizares et al. (1987), i.e.,

$$n_{\rm e}(0) = 0.061 \left( \frac{L_{\rm x}}{10^{41}\ {\rm erg\ s}^{-1}} \right)^{1/2} \left( \frac{a_{\rm x}}{1\ {\rm kpc}} \right)^{-3/2}\ {\rm cm}^{-3}, \qquad (7)$$

we compute electron density profiles according to Eq. (6) for the galaxies in Table 1 which are detected by *EINSTEIN*. We adopted a core radius of 1 kpc for the galaxies for which no determination of the density profile has been reported; here we made an exception for NGC 5044 which shows a very extended surface brightness profile, quite similar to that of NGC 4472 (see Fabbiano et al. 1992). To derive the core radius for NGC 5044, we therefore scaled up the core radius of NGC 4472 (which is 1.5 kpc, Forman et al. 1985) according to the difference in distance. Thus, we adopt $a_x = 3.6$ kpc for NGC 5044.

Laboratory data have shown that the mean free path of keV electrons in solids is a few hundred Å (Seah & Dench 1979), of the same order of magnitude as the size of the dust grains. Thus, most energy of the hot electrons will be deposited in the dust grains. Having derived the electron density profile, we thus calculate the rate of heating of dust by hot electrons $H_e(r)$,

$$H_e(r) = F_e(r)\, E_e\, \pi a^2, \qquad (8)$$

where $E_e = \frac{3}{2}\, kT_e$ is the energy of an electron and

$$F_e(r) = n_e(r) \left( \frac{8\, kT_e}{\pi\, m_e} \right)^{1/2}$$

is the flux of hot electrons at galactic radius $r$ (Spitzer 1978).

*Heating by X-rays*
From the observed X-ray (0.5–3.5 keV) luminosities of the hot gas listed in Table 1 and the core radii mentioned above we derive X-ray intensities $I_x$ (unit: erg cm$^{-2}$ s$^{-1}$ ster$^{-1}$) in the core region, resulting in a heating rate

$$H_x = 4\pi\, I_x\, \pi a^2. \qquad (9)$$

For a typical X-ray luminosity of $10^{41}$ erg s$^{-1}$ and a core radius of 2 kpc we obtain a heating rate by X-rays that is at least two orders of magnitude smaller than the heating rate by optical photons for all the galaxies in our sample. The heating of dust by X-rays can thus safely be neglected in elliptical galaxies (cf. also de Jong et al. 1990).

Dust temperatures are derived by equating the heating rate to the cooling rate of a dust grain by infrared emission

$$L_d = 6.0\, 10^{-8}\, \kappa_0\, m_d\, T_d^5 \text{ erg s}^{-1} \qquad (10)$$

(de Jong et al. 1990) where $m_d = (4\pi/3)\, a^3\, \rho_d$ is the mass of a dust grain. We adopt $a = 0.1$ μm, $\rho_d = 3$ g cm$^{-3}$ and the grain opacity given by Hildebrand (1983), i.e., $\kappa_\nu = \kappa_0\, \lambda_\mu^{-1}$ cm$^2$ g$^{-1}$ with $\kappa_0 = 2.4\, 10^3$, where $\lambda_\mu$ is the wavelength in μm.

The resulting radial distributions of the calculated dust temperature for galaxies in the "RSA sample" are available upon request in tabular form from PG. These tables list the radial profiles of the local dust temperature $T_d$ and the average dust temperature $<T_d>$ inside a given galactic radius $r$ in arcsec, i.e.,

$$<T_d>(r) = \frac{\int_0^r T_d(r')\, 2\pi r'\, dr'}{\pi r^2}. \qquad (11)$$

Here $r$ is the equivalent radius of the ellipse,

$$r \equiv \sqrt{ab} = a\sqrt{1-\epsilon},$$

where a and b are the semimajor and semiminor axes of the elliptical isophote of the galaxy, and $\epsilon$ its ellipticity. The temperatures are tabulated for both heating by optical photons and by hot electrons if appropriate. Examples of the temperature distribution of local and average temperature versus the equivalent radius are shown in Figs. 3 – 5 for the galaxies in the RSA sample which are discussed in Sect. 4.2 below.

In general, the dust temperatures as derived from the *IRAS* data are found to be compatible with those calculated for a diffuse distribution of dust within the inner few kpc.

### 4.2. Properties of the diffusely distributed dust component

In this Section we investigate whether the dust optical depths of the postulated diffuse component of dust in elliptical galaxies as derived from the WTC model are consistent with the observed *IRAS* flux densities. To this end, we derive the mass and infrared luminosity of the diffusely distributed component of dust and compare the dust temperature implied through application of the model of WTC and the radial temperature profiles derived above with the temperature derived directly from the *IRAS* flux densities (corrected for the contribution of circumstellar dust). We limit ourselves to the elliptical galaxies in the RSA sample which are detected at $\geq 2\,\sigma$ at both the 60 and 100 μm wavelengths (cf. Knapp et al. 1989).

Before deriving the temperature and luminosity of the postulated diffuse component of dust, one has to estimate the contribution of the optically visible dust component (if present) to the *IRAS* flux densities. In case of irregular (patchy) dust, this was done by extracting independent, rectangular box-shaped pieces of dust features, comparable in size with the seeing disk. Temperatures, and hence flux densities (cf. Eq. (2)), were calculated for these elements of dust features according to their average (projected) distance from the centre of the galaxy. Apparently regular dust lanes were assumed to be circular disks of uniform density, reaching down to the centre of the galaxy. From Eq. (2), the *IRAS* emission from such a disky dust lane is

$$S_\nu = 9.7\, 10^4\, D^{-2}\, \lambda_\mu^{-4}\, \frac{M_{d,\,opt}}{\pi R_{opt}^2}$$
$$\times \int_0^{R_{opt}} \frac{2\pi r}{\exp(1.44\, 10^4/\lambda_\mu\, T_d(r)) - 1}\, dr \qquad (12)$$

where $S_\nu$ is the *IRAS* flux density in mJy, $D$ is the distance of the Galaxy in Mpc, $M_{d,\,opt}$ is the mass of the dust lane in solar units, $R_{opt}$ is the radial extent of the dust lane in arcsec, $\lambda_\mu$ is the wavelength in μm, and $T_d(r)$ is the radial temperature distribution of the galaxy (cf. Sect. 4.1). The resulting values of the average dust temperatures of dust lanes and/or patches (designated $<T_{d,\,lane}>$) and their contribution to the *IRAS* flux densities (designated $S(60)_{lane}$ and $S(100)_{lane}$, respectively) are listed in Table 2 for the galaxies in our sample.

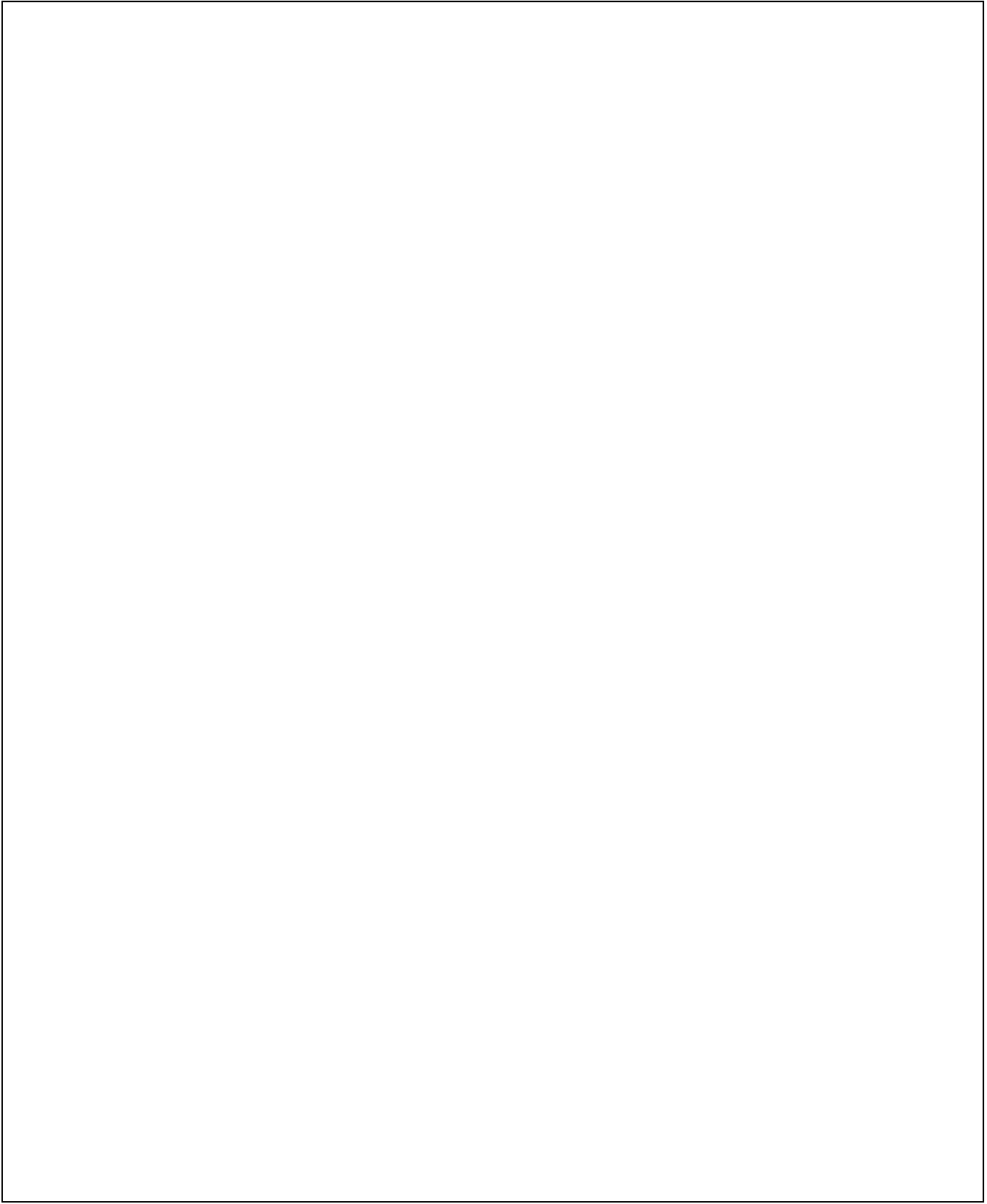

**Fig. 3.** Plots of the local dust temperature due to optical heating (solid line), the average dust temperature due to optical heating (dashed line), and the average dust temperature due to heating by "hot" electrons in X-ray-emitting gas (dotted line) versus galactic radius in kpc for the elliptical galaxies NGC 1395, NGC 1407, NGC 2974, NGC 3377, NGC 3557, NGC 3706, NGC 3904, and NGC 4125. A mixture of graphite and "dirty silicate" (Jones & Merrill 1976) grains of radius 0.1 $\mu$m is assumed

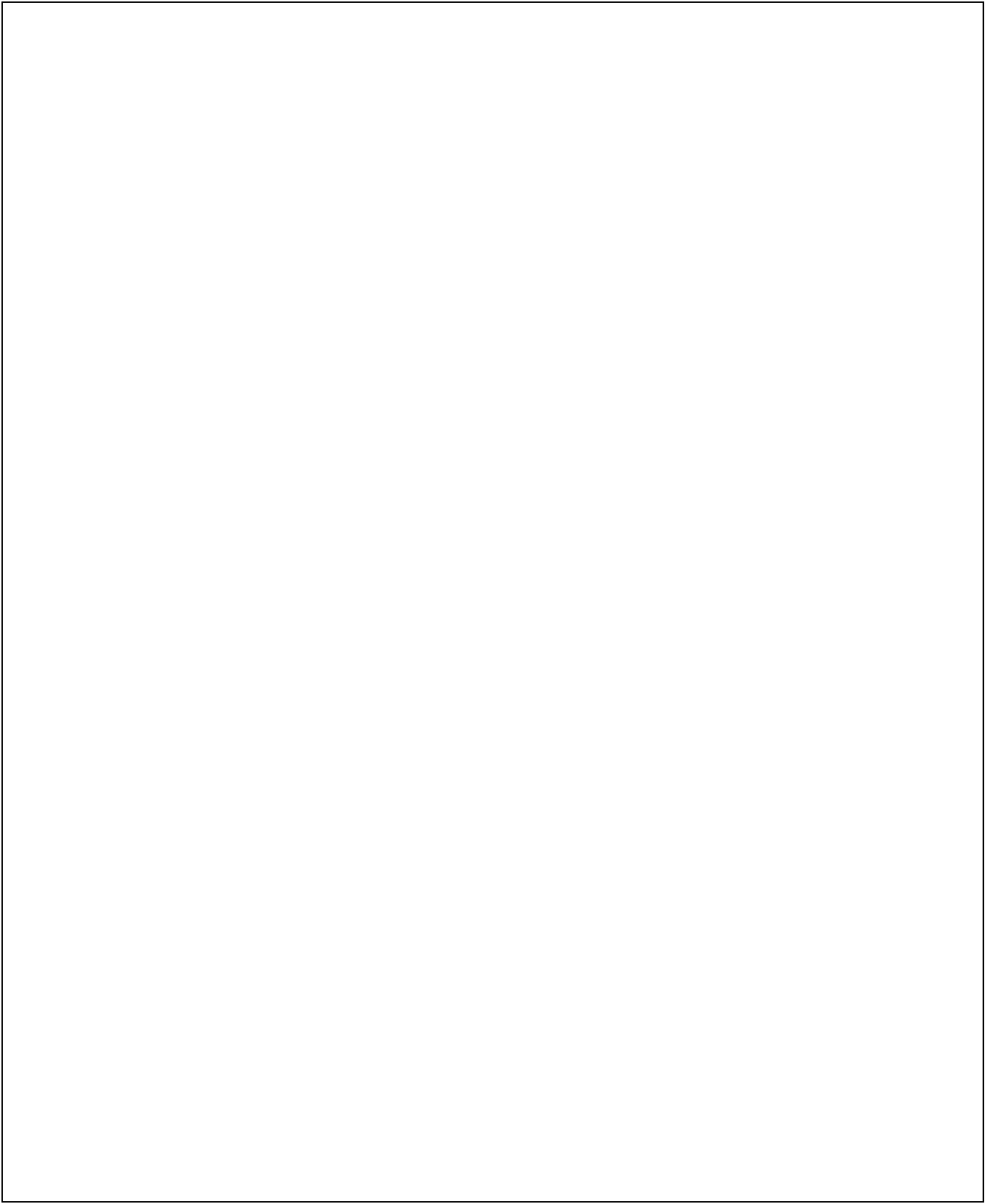

**Fig. 4.** As Fig. 3, for NGC 4261, NGC 4278, NGC 4374, IC 3370, NGC 4486, NGC 4589, NGC 4697, and NGC 4696

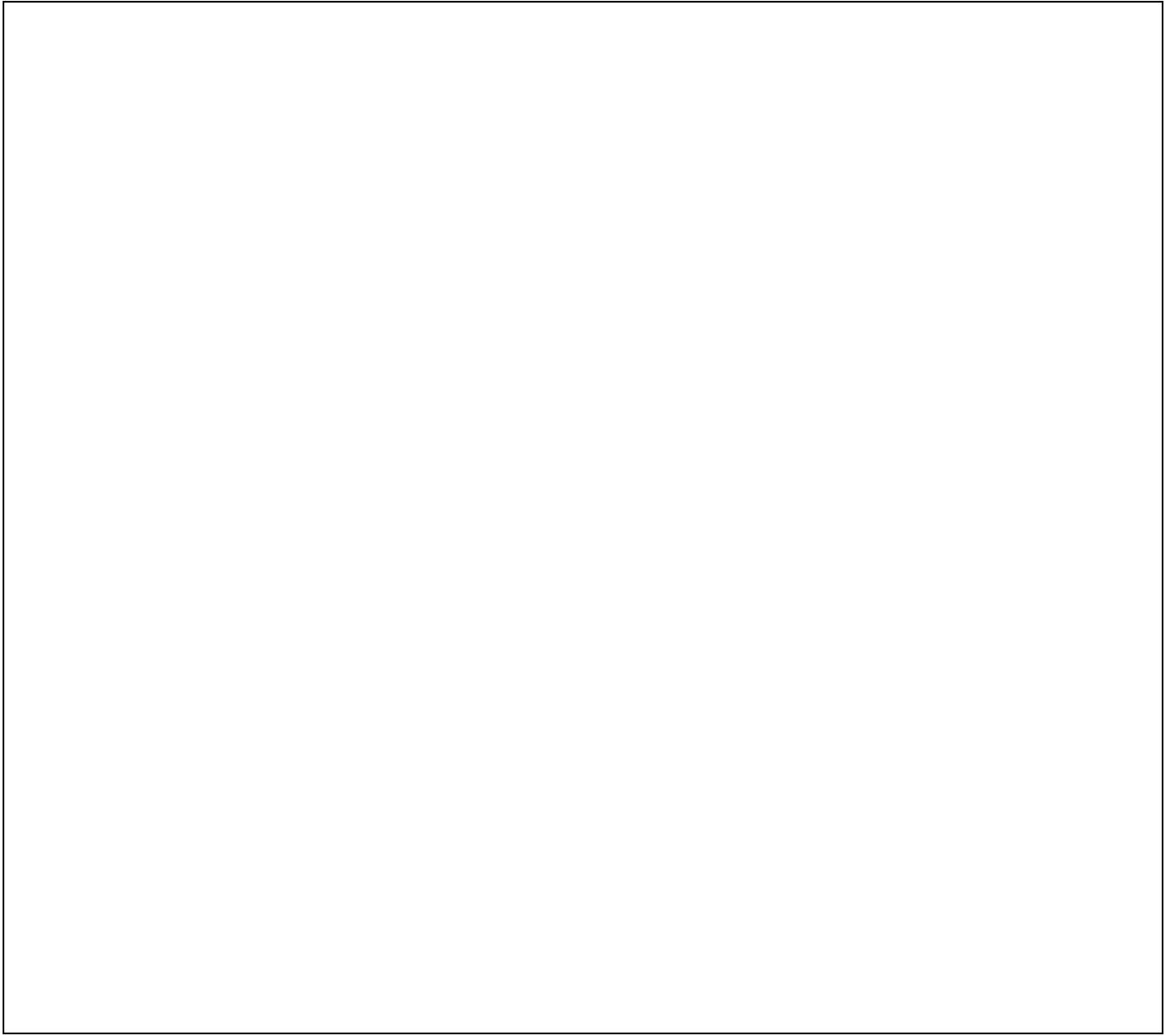

**Fig. 5.** As Fig. 3, for NGC 5018, NGC 5044, IC 4296, NGC 5322, NGC 7144, and IC 1459

The derivation of the contribution of the diffusely distributed component of dust in these galaxies to the *IRAS* flux densities using our derived radial temperature distributions (cf. Sect. 4.1) was done as follows.

1. First one has to subtract the contribution of the optically visible component of dust (if present) to the *IRAS* flux densities. The resulting flux densities were assigned to the postulated diffuse dust component; for the ellipticals in which no optical evidence for the presence of dust was found, the *total IRAS* flux densities were assigned to the diffuse dust component.
2. Dust masses (designated $M_{\rm d,glob}$; column (6) of Table 2) and infrared luminosities of the diffusely distributed dust component were calculated according to Eqs. (2) and (3), respectively.
3. From the $L_{IR}/L_B$ ratio of the diffuse dust component, radial colour gradients and, by implication, total optical depths in the V band were derived according to the model of WTC (cf. Fig. 2). Dust mass column densities were subsequently computed for the diffuse dust component from the total optical depths $\tau_V$ (cf. Paper III).
4. Dividing the implied masses of the diffuse dust component by the dust column densities, outer equivalent radii (designated $r_{\rm glob}$; column (8) of Table 2) of the postulated diffuse dust component were derived for each galaxy. Typical values for the outer radii were of order 2 kpc.

5. *IRAS* flux densities at 60 and 100 $\mu$m for the diffuse dust component (columns (10) and (11) of Table 2) were constructed from the masses of the diffuse dust component by integrating over spheres,

$$S_\nu = 9.7\,10^4\,D^{-2}\,\lambda_\mu^{-4}\,\frac{M_{\rm d,\,glob}}{\frac{4}{3}\pi R_{\rm glob}^3}$$

$$\times \int_0^{R_{\rm glob}} \frac{4\pi r^2}{\exp(1.44\,10^4/\lambda_\mu\,T_{\rm d}(r))-1}\,{\rm d}r$$

where the symbols are the same as in Eq. (12) (but now for a sphere containing $M_{\rm d,\,glob}$ of dust within an outer radius $R_{\rm glob}$). Finally, the average temperatures at the outer radii (designated $<T_{\rm d,\,glob}>$; column (9) of Table 2) were calculated from these *IRAS* flux densities.

A comparison of the observed and reconstructed *IRAS* flux densities (cf. Table 2) reveals that the *observed IRAS* flux densities can in *virtually all elliptical galaxies in the RSA sample* be reproduced *within the 1 $\sigma$ uncertainties* by assuming two components of dust in elliptical galaxies: an optically visible component in the form of dust lanes and/or patches, and a newly postulated dust component which is diffusely distributed within the inner few kpc from the centre of the galaxies.

Heating of the dust by optical photons is found to be generally sufficient to account for the observed dust temperatures as derived from the *IRAS* flux densities, in contrast to conclusions drawn by previous studies of cool interstellar matter in elliptical galaxies (e.g., Brosch 1988; Jura 1986). There are a few exceptions in this respect: NGC 4486 and NGC 5044, both of which have relatively low optical surface brightnesses and thus less effective optical heating. However, both galaxies have bright X-ray coronae. In the case of NGC 5044, the observed dust temperature $T_{\rm d}$ = 48 K is consistent with heating by hot electrons ($n_{\rm e}$ = 8.6 $10^{-2}$ cm$^{-3}$, $T_{\rm e}$ = 1.4 $10^7$ K, cf. Fabbiano et al. 1992) if the typical grain size is $\sim$ 400 Å. Likewise, the dust temperature in NGC 4486 is 49 K, which can only be accounted for by heating by either optical photons or hot electrons ($n_{\rm e}$ = 2 $10^{-2}$ cm$^{-3}$, $T_{\rm e}$ = 2.9 $10^7$ K, $a_{\rm x}$ = 21 kpc, cf. Fabricant et al. 1980) if the typical size of dust grains is $\sim$ 250 Å.

We remind the reader that we have only considered dust which radiates at wavelengths covered by *IRAS*, i.e., with temperatures $\gtrsim$ 25 K. In reality, the postulated diffuse component of dust in elliptical galaxies may generally be expected to extend further out than a few kpc. Future observations with the *Infrared Space Observatory (ISO)* of the RSA sample of elliptical galaxies are foreseen, and may reveal this cooler dust component in elliptical galaxies.

## 5. Origin of the different components of dust

As to the origin of cool interstellar matter in elliptical galaxies, a consensus is developing that dust in these systems *always* has an external origin, i.e., accreted during a galaxy merger or interaction. For instance, spectroscopic studies of the velocity fields of gas and stars in elliptical galaxies with dust lanes show that the gas which is associated with the dust lanes is generally dynamically decoupled from the stellar body, i.e., in disks rotating at random orientations with respect to the apparent major axis of the elliptical galaxy (Raimond et al. 1981; Möllenhoff 1982; Sharples et al. 1983; Bertola et al. 1985; Caldwell et al. 1986; Davies & Illingworth 1986; Bertola & Bettoni 1988; Möllenhoff & Bender 1989; Kim 1989; Goudfrooij 1994b). Since it is difficult (if not impossible) to envision a evolutionary scenario for a *primordial* galaxy in which gaseous and stellar components become dynamically decoupled, this strongly suggests an external origin for the dust and gas.

Other important evidence in favour of an external origin of the dust and gas is provided in several cases of X-ray-emitting elliptical and cD galaxies with suspected cooling flows. These galaxies often contain extended regions of ionized gas which have been argued to arise as condensations in thermally instable regions in the cooling flow (see, e.g., Fabian et al. 1991). The lifetime of a dust grain against collisions with hot protons and $\alpha$-particles ("sputtering") in a hot gas with $T_{\rm e} \sim 10^7$ K is

$$\tau_{\rm d} \equiv a \left|\frac{{\rm d}a}{{\rm d}t}\right|^{-1} \simeq 2\,10^5 \left(\frac{n_{\rm H}}{{\rm cm}^{-3}}\right)^{-1} \left(\frac{a}{0.1\,\mu{\rm m}}\right)\,{\rm yr} \qquad (13)$$

(Draine & Salpeter 1979a), which is of order only $10^6 - 10^7$ yr. Hence, the emission-line filaments in these galaxies are expected to be dust-free. This is illustrated by the finding that the intergalactic medium within the Coma cluster is depleted in dust by a factor of $\sim$ 140 with respect to the Galactic dust-to-gas ratio (Dwek et al. 1990). However, dust *has* been found to be quite commonly associated with filaments of ionized gas in these galaxies (Paper II; Jørgensen et al. 1983; Hansen et al. 1985; Sparks et al. 1989; Kim 1989; de Jong et al. 1990; Goudfrooij et al. 1990; Macchetto & Sparks 1991; Goudfrooij 1991; Nørgaard-Nielsen et al. 1993; Sparks et al. 1993). Furthermore, Donahue & Voit (1993) recently concluded in their study of galaxies with suspected cooling flows that the line-emitting filaments in these galaxies contain dust, because the observed upper limit on the intensity of the [Ca II] emission line at 7291 Å indicates that Calcium is depleted onto dust grains.

Although the evidence mentioned above provides a strong case for an external origin of dust in elliptical galaxies, these arguments are all based on the *optically visible component* of dust, which only accounts for $\lesssim$ 10% of the total amount of dust in elliptical galaxies (cf. Sect. 4). So the question arises: What is the origin of the postulated diffusely distributed component of dust?

An example of an elliptical galaxy in which *all* dust probably has an external origin is NGC 4696, the dominant galaxy of the Centaurus cluster. This galaxy contains an infalling, filamentary dust lane in its central region, associated with ionized gas (Sparks et al. 1989; de Jong et al. 1990). De Jong et al. showed that the X-ray, optical, and *IRAS* data of NGC 4696 can be explained in the context of an "evaporation flow" scenario. They suggest that cool clouds of gas and dust have been brought in during the accretion of a gas-rich dwarf galaxy. These cloudlets (with sizes and densities typical of cores of molecular clouds in our own Galaxy) gradually evaporate in the X-ray-emitting gas on a time scale of $10^8$ yr, thereby exposing the dust grains to a hot environment in which they are

**Table 2.** Computed temperatures and *IRAS* flux densities of patchy and diffuse dust components

| Galaxy | log $M_{\rm d, lane}$ [$M_\odot$] | $\langle T_{\rm d, lane}\rangle$ [K] | $S(60)_{\rm lane}$ [mJy] | $S(100)_{\rm lane}$ [mJy] | log $M_{\rm d, glob}$ [$M_\odot$] | $\frac{\Delta(B-I)}{\Delta(\log r)}$ (dust) | $r_{\rm glob}$ [Kpc] | $\langle T_{\rm d, glob}\rangle$ [K] | $S(60)_{\rm glob}$ [mJy] | $S(100)_{\rm glob}$ [mJy] | $S(60)$ [mJy] | $S(100)$ [mJy] |
|---|---|---|---|---|---|---|---|---|---|---|---|---|
| (1) | (2) | (3) | (4) | (5) | (6) | (7) | (8) | (9) | (10) | (11) | (12) | (13) |
| NGC 1395 | — | — | — | — | 5.60 | 0.037 | 2.9 | 29/*34* | 135 | 496 | 50 ± 26 | 300 ± 042 |
| NGC 1407 | — | — | — | — | 5.35 | 0.040 | 2.6 | 31/*35* | 133 | 423 | 140 ± 30 | 430 ± 065 |
| NGC 2974 | 4.67 | 32/*31* | 29 | 86 | 6.17 | 0.162 | 2.7 | 29/*26* | 418 | 1592 | 430 ± 33 | 1690 ± 047 |
| NGC 3377 | 4.00 | 34 | 63 | 162 | 3.85 | 0.040 | 0.4 | 38 | 104 | 191 | 140 ± 46 | 310 ± 057 |
| NGC 3557 | — | — | — | — | 5.94 | 0.070 | 3.2 | 33 | 307 | 796 | 250 ± 47 | 670 ± 150 |
| NGC 3706 | — | — | — | — | 5.27 | 0.045 | 1.7 | 34 | 91 | 224 | 70 ± 48 | 200 ± 068 |
| NGC 3904 | — | — | — | — | 5.21 | 0.086 | 1.2 | 35 | 208 | 500 | 210 ± 49 | 480 ± 160 |
| NGC 4125 | 5.66 | 31 | 189 | 594 | 5.34 | 0.098 | 1.3 | 38 | 355 | 675 | 720 ± 45 | 1480 ± 061 |
| NGC 4261 | — | — | — | — | 4.64 | 0.031 | 1.0 | 38 | 63 | 119 | 80 ± 35 | 130 ± 043 |
| NGC 4278 | 4.38 | 26 | 16 | 80 | 5.18 | 0.106 | 1.1 | 32 | 562 | 1610 | 600 ± 54 | 1650 ± 053 |
| NGC 4374 | 4.44 | 36/*30* | 107 | 234 | 5.03 | 0.054 | 1.2 | 35/*30* | 375 | 811 | 510 ± 27 | 1030 ± 109 |
| IC 3370 | 5.55 | 33 | 91 | 244 | 6.69 | 0.234 | 4.1 | 29 | 512 | 1885 | 570 ± 28 | 2010 ± 106 |
| NGC 4486 | 3.18 | 35/*37* | *7* | *14* | 4.07 | 0.027 | 0.6 | 39/*37* | 77 | 136 | 400 ± 41 | 360 ± 091 |
| NGC 4589 | 4.94 | 30/*24* | 31 | 106 | 5.49 | 0.090 | 1.6 | 32/*25* | 169 | 494 | 210 ± 31 | 590 ± 136 |
| NGC 4697 | — | — | — | — | 5.21 | 0.050 | 1.6 | 34/*27* | 441 | 1107 | 470 ± 23 | 1100 ± 067 |
| NGC 4696 | 5.66 | 25/*24* | 17 | 91 | 6.62 | 0.061 | 7.2 | 24/*24* | 111 | 661 | 100 ± 23 | 740 ± 131 |
| NGC 5018 | 5.46 | 35 | 156 | 362 | 5.78 | 0.221 | 1.5 | 40 | 798 | 1325 | 980 ± 41 | 1650 ± 080 |
| NGC 5044 | 4.38 | 33/*40* | *31* | *52* | 3.62 | 0.053 | 0.7 | 35/*40* | 6 | 10 | 140 ± 57 | 130 ± 065 |
| IC 4296 | — | — | — | — | 5.36 | 0.044 | 2.0 | 35/*37* | 97 | 196 | 140 ± 59 | 230 ± 073 |
| NGC 5322 | — | — | — | — | 5.67 | 0.061 | 2.5 | 34 | 305 | 760 | 430 ± 38 | 890 ± 067 |
| NGC 7144 | — | — | — | — | 5.34 | 0.050 | 1.7 | 30 | 87 | 289 | 90 ± 34 | 290 ± 095 |
| IC 1459 | 5.28 | 31/*30* | 105 | 330 | 5.14 | 0.090 | 1.1 | 39/*30* | 383 | 661 | 520 ± 31 | 1050 ± 091 |

*Notes to Table 2.*
Column (2) lists the dust mass of the optically visible ("patchy") dust component as derived from the optical extinction values, and column (3) lists its average temperature as derived from heating by optical photons [roman font] and from heating by "hot" electrons [*italic* font] (cf. Sect. 4.1). Columns (4) and (5) list the calculated 60 and 100 $\mu$m flux densities for the optically visible dust component, respectively. Column (6) lists the dust mass of the proposed "diffuse" dust component, derived after subtraction of columns (4) and (5) from the *IRAS* flux densities. Column (7) lists the $(B-I)$ colour gradient due to dust alone, as predicted from the $L_{IR}/L_B$ ratio of the ellipticals (cf. Fig. 2). Column (8) lists the galactic equivalent radius at which the integrated mass of the predicted diffuse dust component equals the dust mass in column (6). Column (9) lists the average temperature of the diffuse dust component (cf. column (3)), and columns (10) and (11) list the calculated 60 and 100 $\mu$m flux densities for the diffuse dust component, respectively. The flux densities in columns (4), (5), (10), and (11) are given in *italic* font when heating by hot electrons is found to be more effective than heating by optical photons. Columns (12) and (13) list the *IRAS* flux densities at 60 and 100 $\mu$m, respectively, along with their 1 $\sigma$ uncertainties (taken from Knapp et al. 1989) heated by hot electrons and destroyed by sputtering. Sparks et al. (1989) independently arrived at a similar scenario. Further support for this scenario is provided by the observation that the dusty filaments in NGC 4696 display a normal wavelength dependence of extinction (consistent with that of Galactic dust, cf. Paper III), showing that the grain properties have not (yet) substantially been modified.

From an optical, radio, and X-ray point of view, NGC 4486 (M 87) closely resembles NGC 4696 (cf. Matilsky et al. 1985). Also in NGC 4486, patches of dust have been found to be associated with the "H$\alpha$+[N II]-emitting jet" in deep optical images (Sparks et al. 1993; Paper II). However, the resemblance does not hold in the far-infrared. NGC 4486 has a much higher dust temperature (49 K, instead of 24 K for NGC 4696), which indicates that the typical size of dust grains in NGC 4486 is significantly smaller than the canonical size of Galactic dust (cf. Sect. 4.1). This suggests that the cool cloud component brought in during a merging collision (as proposed for the case of NGC 4696, cf. Sparks et al. 1989; de Jong et al. 1990) is nearly completely evaporated in the hot gas associated with NGC 4486.

The dust in NGC 4486 is probably replenished by condensation of dust grains in the winds and circumstellar shells of late-type stars (see also de Jong et al. 1990). Using the present-day mass loss rate of $1.5\,10^{-11}$ $(L_B/L_\odot)$ $M_\odot$ yr$^{-1}$ (Faber & Gallagher 1976; Knapp et al. 1992), we derive a gas production rate $\dot{M}_{\rm gas} = 0.7$ $M_\odot$ yr$^{-1}$. For a dust destruction timescale $\tau_{\rm d} = 3\,10^6$ yr in NGC 4486 ($a = 250$ Å, $n_{\rm e} = 2\,10^{-2}$ cm$^{-3}$, $n_{\rm H} = 0.83\,n_{\rm e}$, $T_{\rm e} = 2.9\,10^7$ K), this results exactly in the observed dust mass of $1.5\,10^4$ $M_\odot$ if the gas-to-dust ratio in the region of dust condensation in the winds and circumstellar shells of late-type stars has the plausible value of $\sim 150$. Hence, the dust observed in NGC 4486 is probably mainly produced by stellar mass loss and destroyed by sputtering in hot gas.

This result can be used as a tool to further constrain the origin of dust in elliptical galaxies. In Fig. 6 we show the relationship of the dust mass derived from the *IRAS* data with the blue luminosities of the galaxies in the RSA sample. The dashed lines in Fig. 6 connect the loci where all dust could have an internal origin, i.e., replenished by stellar mass loss at the rate given by Faber & Gallagher (1976) and destroyed by (e.g.) sputtering, for different values of the destruction time scale (cf. Eq. (13)). We adopted a gas-to-dust ratio of 100. The line with $\tau_{\rm d} = 10^{7.5}$ yr should be regarded as a maximum destruction time scale in ellipticals known to contain hot, X-ray-emitting gas, since the electron density exceeds $5\,10^{-3}$ cm$^{-3}$ in the inner $\sim 10$ kpc of all galaxies studied in sufficient detail (cf. Matilsky et al. 1985; Forman et al. 1985; Trinchieri et al. 1986; Canizares et al. 1987). As Fig. 6 shows, most elliptical galaxies in the RSA sample in which X-ray emission has been detected contain more dust than can be accounted for by stellar mass loss alone. This favours the view that the dust in elliptical galaxies generally has an external origin.

As can be seen in Fig. 6, we confirm the absence of a correlation between the mass of dust and the blue luminosity of elliptical galaxies (cf. Forbes 1991). On first sight, this indicates

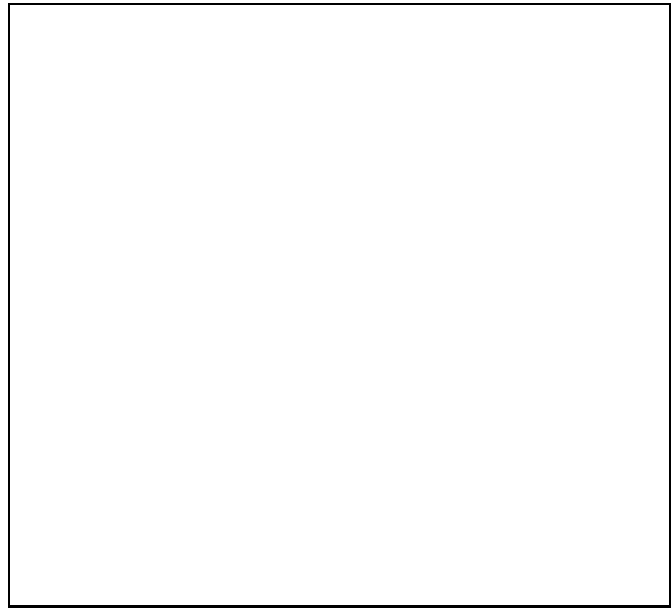

**Fig. 6.** The relationship of the mass of cool dust (as derived from the *IRAS* 60 and 100 $\mu$m flux densities) and the blue luminosity of elliptical galaxies in the RSA sample, as tabulated in Table 1. Filled symbols represent galaxies with optically visible dust lanes and/or patches, whereas open symbols represent galaxies in which no sign of optical extinction has been found. Triangles pointing up- or downwards represent galaxies which are detected by *IRAS* at either 100 or 60 $\mu$m only, respectively. Encircled symbols represent galaxies with detected X-ray emission from hot gas. The position of M 87 is marked. The dashed lines connect loci where dust is replenished by stellar mass loss, and destroyed by sputtering at the given time scale (in dex), as discussed in the text

that there is no causal relationship between the dust content and the present-day population of stars within elliptical galaxies, in contrast with the situation in spiral galaxies (e.g., Young et al. 1989), which would again suggest that dust in elliptical galaxies has an external origin (see also Knapp et al. 1985; Forbes 1991; Lees et al. 1991). However, this conclusion is somewhat premature since the typical dust destruction timescale in elliptical galaxies is expected to be significantly different among individual objects. Among the various destruction mechanisms for dust grains in the interstellar medium, sputtering in hot, X-ray-emitting gas is by far the most effective one (Spitzer 1978; Draine & Salpeter 1979a, b; Draine & Anderson 1985; Dwek 1986; Dwek et al. 1990; Paper III). The fate of dust within elliptical galaxies is therefore mainly determined by the presence of hot, X-ray-emitting coronal gas and its physical properties (i.e., radial temperature and density profiles). However, the present status of knowledge of these important matters is quite limited (cf. Sect. 4.1), since e.g., most of the galaxies in the RSA sample have not been observed by the *EINSTEIN* satellite. Future analysis of the *ROSAT* all-sky survey data and, especially, the gain in sensitivity and spectral resolution of the *ASCA* satellite should be very valuable in determining typical destruction timescales for dust in elliptical galaxies.

An alternative explanation for the finding that the dust content of elliptical galaxies containing hot, X-ray-emitting gas is higher than can be accounted for by mass loss of stars within the galaxies might be that dust is being produced in association with low-mass star formation like Hansen et al. (1994) recently proposed for the case of Hydra A. The evidence to date for star formation in cooling flow galaxies amounts to total star formation rates that are only a very small fraction (usually $< 1\%$) of the mass deposition rates that have been deduced from X-ray observations (up to $\dot{M} \sim 400\, M_\odot\, \mathrm{yr}^{-1}$; see, e.g., Fabian et al. 1991; McNamara & O'Connell 1992; Hansen et al. 1994). This is usually interpreted in terms of an initial mass function of stars formed in cooling flows which is heavily biased towards "invisible" low-mass stars. However, one would expect that even in the case of low-mass star formation, the cooling gas should pass through a molecular phase, whereas the sensitive CO measurements of Braine & Dupraz (1994) did not result in detections for any of the 8 dominant cluster galaxies with reported mass deposition rates $\dot{M} > 100\, M_\odot\, \mathrm{yr}^{-1}$ in their sample. On the basis of these non-detections, Braine & Dupraz rule out the high mass deposition rates as determined from the "standard model" of cooling flows. Thus, we are tempted to conclude that the evidence for substantial rates of star formation in X-ray-emitting elliptical galaxies needs further confirmation, especially for the X-ray-emitting galaxies in our "RSA sample", for which no evidence for young stellar populations has been found (cf. Papers I, II, and III).

## 6. Concluding remarks

We have investigated *IRAS* far-infrared observations of a complete, blue magnitude limited sample of 56 elliptical galaxies selected from the Revised Shapley-Ames Catalog. The luminosity range of the galaxies in this sample is $-19.5 \gtrsim M_{B_\mathrm{T}} \gtrsim -23.0$. Data from a homogeneous optical CCD imaging survey as well as published X-ray data from the *EINSTEIN* satellite are used to constrain the *IRAS* data. The main conclusions from this paper can be summarized as follows.

1. Dust masses derived from the emission in the 60 and 100 $\mu$m *IRAS* passbands are roughly an order of magnitude *higher* than those estimated from optical extinction, in strong contrast with the situation in spiral galaxies. We argue that the "missing dust" is diffusely distributed over the inner few kpc of the elliptical galaxies.
2. Using *observed* radial optical surface brightness profiles, we have systematically investigated possible heating mechanisms for the dust within elliptical galaxies. The observed far-infrared colour temperatures are found to be generally consistent with heating of dust by the interstellar radiation field and by hot electrons in X-ray-emitting gas (if appropriate).
3. Employing model calculations which involve the transfer of stellar radiation in a spherical distribution of stars mixed with a diffuse distribution of dust, we show that the observed infrared luminosities of the postulated diffusely distributed dust component in elliptical galaxies imply total dust optical depths in the range $0.1 \lesssim \tau_V \lesssim 0.7$ and radial colour gradients $0.03 \lesssim (\Delta(B-I)/\Delta\log r) \lesssim 0.25$, consistent with observed colour gradients. We argue that the effect of dust extinction should be taken seriously in the interpretation of colour gradients in elliptical galaxies.
4. We show that the amount of dust observed in elliptical galaxies is generally higher than that expected from production by mass loss of stars within elliptical galaxies and destruction by sputtering in hot gas. This suggests that most of the dust in elliptical galaxies generally has an external origin.

*Acknowledgements.* We thank the referee and S. Pellegrini for giving constructive comments on the manuscript. We have made use of the NASA/IPAC Extragalactic Database (NED) which is operated by the Jet Propulsion Laboratory, Caltech, under contract with the National Aeronautics and Space Administration.